\documentclass[aps,prb,twocolumn,superscriptaddress]{revtex4}%
\usepackage{amsmath}
\usepackage{bm}
\usepackage{graphicx}
\usepackage{amsfonts}
\usepackage{amssymb}
\begin{document}
\title{ Kondo and Dicke  effect in quantum-dots side coupled to a quantum wire}
\author{Pedro A. Orellana}
\address{Departamento de F\'{\i}sica,
Universidad Cat\'{o}lica del Norte, Casilla 1280, Antofagasta,
Chile}
\author{Gustavo A. Lara}
\address{Departamento de F\'{\i}sica,
Universidad de Antofagasta, Casilla 170, Antofagasta, Chile}
\author{Enrique V. Anda}
\address{Departamento de F\'{\i }sica,
P. U. Cat\'{o}lica do Rio de Janeiro, C.P. 38071-970, Rio de
Janeiro, RJ, Brazil}


\begin{abstract}
Electron tunneling through quantum-dots side coupled to a quantum
wire, in equilibrium and nonequilibrium Kondo regime, is studied.
The mean-field finite-$U$ slave-boson formalism is used to obtain
the solution of the problem. We have found that the transmission
spectrum shows a structure with two anti-resonances localized at the
renormalized energies of the quantum dots. The DOS of the system
shows that when the Kondo correlations are dominant there are two
Kondo regimes with its own Kondo temperature. The above behavior of
the DOS can be explained by quantum interference in the transmission
through the two different resonance states of the quantum dots
coupled to common leads. This result is analogous to the Dicke
effect in optics. We investigate the many body Kondo states as a
function of the parameters of the system.

\end{abstract}

\maketitle

\section{Introduction}

The Kondo effect in quantum dots (QDs) has been extensively
studied in the last years \cite{GR88,GSMAMK98,Kouwenhoven}. The
QDs allow studying systematically the quantum-coherence many-body
Kondo state, due to the possibility of continuous tuning the
relevant parameters governing the properties of this state, in
equilibrium and nonequilibrium situations. Recently Kondo effect
has been studied in side attach quantum dot \cite{THCP02} and
parallel quantum dots \cite{tanaka,sakano}. Recent electron
transport experiments showed that Kondo and Fano resonances occur
simultaneously \cite{GGHK00}. Multiple scattering of traveling
electronic waves on a localized magnetic state are crucial for the
formation of both resonances. The condition for the Fano resonance
is the existence of two scattering channels: a discrete level and
a broad continuum band \cite{fano}.

An alternative configuration consists of two single QDs side
attached to a perfect quantum wire (QW). This structure is
reminiscent of the cross-bar-shaped quantum wave guides
\cite{DRVRPM00}. In this case, the QDs act as scattering centers
in close analogy with the traditional Kondo effect \cite{KCKS01}.
This configuration was study previously by  Stefa\'nski
\cite{stefanski} and Tamura et at. \cite{glazman}.

In this work we study the transport properties of two single quantum
dots side coupled to a quantum wire in the Kondo regime. We use the
finite-$U$ slave boson mean-field approach, which was initially
developed by Kotliar and Ruckenstein \cite{KR86} and used later by
Bing Dong and X. L. Lei to study the transport through coupled
double quantum dots connected to leads \cite{DL01}.This approach
enforces the correspondence between the impurity fermions and the
auxiliary bosons to a mean-field level to release the $U=\infty$
restriction. In quantum dots, this approach allows to treat the
dot-lead coupling nonperturbatively for an arbitrary strength of the
Coulomb interaction $U$ \cite{DL01}. We have found that the
transmission spectrum shows a structure with two anti-resonances
localized at the renormalized energies of the quantum dots. The DOS
of the system shows that when the Kondo correlations are dominant
there are two Kondo regimes each with its own Kondo temperature. The
above behavior of the DOS can be explained by quantum interference
in the transmission through the two different resonance states of
the quantum dots coupled to common lead. This phenomenon is in
analogy to  the Dicke effect in quantum optics, that takes place in
the spontaneous emission of two closely-lying atoms radiating a
photon into the same environment~\cite{dicke}. In the electronic
case, however, the decay rates (level broadening) are produced by
the indirect coupling of the up-down QDs, giving rise to a fast
(\emph{superradiant\/}) and a slow (\emph{subradiant\/}) mode.
Recently, Brandes reviewed the Dicke effect in mesoscopic
systems~\cite{brandes}.


\section{Model}

\begin{figure}[h]
\centering
  \begin{picture}(220,100)(0,0)
    \thicklines
    \put(40,50){\line(1,0){140}}
    \put(110,50){\line(0,-1){30}}
    \put(110,50){\line(0,1){30}}
    \put(110,80){\circle*{12}}
    \put(110,20){\circle*{12}}
    \put(100,90){\makebox(0,0){$\varepsilon_{1}$}}
    \put(100,10){\makebox(0,0){$\varepsilon_{2}$}}
    \put(120,30){\makebox(0,0){$t_{0}$}}
    \put(120,65){\makebox(0,0){$t_{0}$}}
    \put(170,55){\makebox(0,0)[b]{$\varepsilon_{w}=0$}}
    \put(50,70){\makebox(0,0)[b]{QW}}
    \put(10,50){\makebox(0,0)[r]{$\mu_{L}$}}
    \put(210,50){\makebox(0,0)[l]{$\mu_{R}$}}
    \put(40,50){\makebox(0,0)[r]{$.\, .\, .\, .\, .$}}
    \put(180,50){\makebox(0,0)[l]{$.\, .\, .\, .\, .$}}
    \multiput(50,50)(20,0){7}{\circle*{4}}
    \qbezier(0,30)(30,50)(0,70)
    \qbezier(220,30)(180,50)(220,70)
  \end{picture}
  \caption{Scheme of side-coupled quantum dots attached
           laterally to a perfect quantum wire (QW). The QW is
           coupled to the left ($L$) and right ($R$) noninteracting
           leads.}
  \label{f:esquema}
\end{figure}
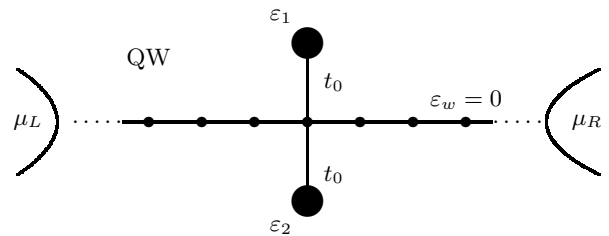

Let us consider two single quantum dot (2QD) side coupled to a
perfect quantum wire (QW) (see Fig.~\ref{f:esquema}). We adopt the
two-impurities Anderson Hamiltonian. Each dot has a single level
energy $\varepsilon_{l}$ (with $l = 1, 2$), and a intra-dot Coulomb
repulsion $U$. The two side attached quantum-dots are coupled to the
QW with coupling $t_{0}$. The QW sites have local energies
$\varepsilon_{wi,\sigma} =0$  and a hopping parameter $t$.

The corresponding Hamiltonian model is,

\begin{widetext}
\begin{equation}
H_{0} = -t \sum_{i,\sigma} \left( c_{i,\sigma}^{\dag} c_{i+1, \sigma}
                             + \text{H.c.} \right)
+ \sum_{l=1,2,\sigma} \left[
-t_{l,\sigma}\left( c_{0,\sigma}^{\dag} f_{l,\sigma}      + \text{H.c.} \right)
+ \left( \varepsilon_{l,\sigma}+\frac{U}{2}\hat{n}_{l,-\sigma}\right) \hat{n}_{l,\sigma}
      \right] \,
\end{equation}
\end{widetext}

\noindent where $c_{i,\sigma}^{\dag}$ ($c_{i,\sigma}$) is the
creation (annihilation) operator of an electron with spin $\sigma$
at the $i$-th site of the quantum wire; $f_{l,\sigma}^{\dag}$
($f_{l,\sigma}$) is the creation (annihilation) operator of an
electron with spin $\sigma$ in the l-th QD and $\hat{n}_{l,\sigma}$
is the corresponding number operator.

To find the solution of this correlated fermions system, we appeal
to an analytical approach where, generalizing the infinite-$U$
slave-boson approximation \cite{C84} the Hilbert space is enlarged
at each site, to contain in addition to the original fermions a set
of four bosons \cite{KR86} represented by the creation
(annihilation) operators $e_{l}^{\dag}$ ($e_{l}$),
$p_{l,\sigma}^{\dag}$ ($p_{l,\sigma}$), and $d_{l}^{\dag}$ ($d_{l}$)
for the $l$-th dot.They act as projectors onto empty, single
occupied (with spin up and down) and doubly occupied electron
states, respectively. Then, each creation (annihilation) operator of
an electron with spin $\sigma$ in the l-th QD, is substituted by $
f_{l,\sigma}^{\dag} \tilde{Z}_{l,\sigma}^{\dag}$
($\tilde{Z}_{l,\sigma} f_{l,\sigma}$) where:
\begin{equation}
\tilde{Z}_{l,\sigma} =
  \frac{e_{l}^{\dag} p_{l, \sigma} + p_{l, -\sigma}^{\dag} d_{l}}
       {\sqrt{1-d_{l}^{\dag}d_{l}-p_{l,\sigma}^{\dag} p_{l,\sigma}}
        \sqrt{1-e_{l}^{\dag}e_{l}-p_{l,-\sigma}^{\dag}p_{l,-\sigma}} }
\end{equation}
\noindent The denominator is chosen to reproduce the correct
$U\rightarrow0$ limit in the mean-field approximation without
changing neither the eigenvalues nor the eigenvector.

The constraint, i.e., the completeness relation
$\sum_{\sigma}p_{l,\sigma}^{\dag}
p_{l,\sigma}+b_{l}^{\dag}b_{l}+d_{l}^{\dag}d_{l}=1$ and the
condition among fermions and bosons
$n_{l,\sigma}-p_{l,\sigma}^{\dag}p_{l,\sigma}-d_{l}^{\dag}d_{l} =
0$, will be  incorporated with Lagrange multipliers
$\lambda_{l}^{(1)}$ and $\lambda_{l,\sigma}^{(2)}$ into the
Hamiltonian. Also in the mean-field approximation all the boson
operators are replaced by their expectation value
which can be chosen, without loss of generality, as real numbers.

The Hamiltonian in this new and enlarged Hilbert space, is,  ${\cal H}=H_{b}+H_{e}$,
where

\begin{eqnarray}
H_{b} &=& \sum_{l=1,2} \lambda_{l}^{(1)} \left( p_{l,\uparrow}^{2} +
        p_{l,\downarrow}^{2} + e_{l}^{2} + d_{l}^{2} - 1 \right) \nonumber   \\
      & & - \sum_{l=1,2,\sigma}  \lambda_{l,\sigma}^{(2)} \left( p_{l, \sigma}^{2} +
d_{l}^{2} \right)+ U\sum_{l=1,2} d_{l}^{2}  \, ,
\label{h_boson}
\end{eqnarray}

\noindent depends explicitly only upon the boson expectation
values and the Lagrange multipliers, and

\begin{eqnarray}
H_{e} &=& -t \sum_{i,\sigma} \left( c_{i,\sigma}^{\dag} c_{i+1, \sigma}
                               + \text{H.c.} \right)  \nonumber \\
      & & + \sum_{l=1,2,\sigma} \left[ -\tilde{t}_{l,\sigma}
          \left( c_{0,\sigma}^{\dag} f_{l,\sigma} + \text{H.c.} \right)
            +\tilde{\varepsilon}_{l, \sigma} n_{l,\sigma}    \right]
\label{h_electron}
\end{eqnarray}

\noindent is a tight-binding Hamiltonian that depends implicitly
on the boson expectation values through the parameters:
$\tilde{\varepsilon}_{l,\sigma}=\varepsilon_{l,\sigma}+\lambda_{l,\sigma}^{(2)}$,
  $\, \tilde{t}_{l,\sigma} = t_{0}\langle\tilde{Z}_{l,\sigma}\rangle$.

 As we work at zero temperature, the boson operators expectation values and
the Lagrange multipliers are determined by minimizing the energy
$\langle {\cal H} \rangle $ with respect to these quantities. It
is obtained in this way, a set of nonlinear equations for each
quantum dot, relating the expectation values of the four bosonic
operators, the three Lagrange multipliers and the electronic
expectation values,

\begin{subequations}
\label{e:minimizing}%
\begin{align}
p_{l,\sigma }^{2}  & = \langle \hat{n}_{l,\sigma} \rangle -d_{l}^{2} \,,\\
e_{l}^{2}  & = 1- \sum_{s} \langle \hat{n}_{l,s}\rangle +d_{l}^{2} \,,\\
\lambda_{l}^{\left( 1\right) }  & =
    \frac{t_{0}}{e_{l}}
        \sum_{s} \langle f_{l,s}^{\dag } c_{0,s } \rangle
              \frac{\partial \langle \tilde{Z}_{l,s}\rangle }{\partial e_{l}}  \,,\\
\lambda_{l}^{\left( 1\right) }-\lambda_{l,\sigma}^{\left( 2\right) }   & =         \frac{t_{0}}{p_{l,\sigma}}
        \sum_{s} \langle f_{l,s}^{\dag } c_{0,s } \rangle
              \frac{\partial \langle \tilde{Z}_{l,s}\rangle }{\partial p_{l,\sigma}}  \,,\\
U+\lambda_{l}^{\left( 1\right) }-\sum_{s}\lambda_{l,s}^{\left( 2\right) }   & =
    \frac{t_{0}}{d_{l}}
        \sum_{s} \langle f_{l,s}^{\dag } c_{0,s } \rangle
              \frac{\partial \langle \tilde{Z}_{l,s}\rangle }{\partial d_{l}}  \,.
\end{align}
\end{subequations}

\noindent  where $l$ is the dot index, $s$, $\sigma $ are spin
indexes and $\langle \tilde{Z}_{l,s}\rangle$ satisfies,

\begin{equation}
\langle \tilde{Z}_{l,s}\rangle = \frac {  p_{l,s}\left( e_{l}+d_{l}
\right)  } {  \sqrt{\left( 1-d_{l}^2-p_{l,s}^2 \right) \left(
1-e_{l}^2-p_{l,-s}^2 \right) }  }.
\end{equation}

To obtain the electronic expectation values $\langle \cdots \rangle $, the Hamiltonian, $H_{e}$, is diagonalized and their stationary states
can be written as
\begin{equation}
\left\vert \psi_{k}\right\rangle =\sum_{j=-\infty}^{\infty}a_{j}^{k}\left\vert
j\right\rangle +\sum_{l=1}^{2}b_{l}^{k}\left\vert l\right\rangle \,,
\end{equation}
where $a_{j}^{k}$ and $b_{l}^{k}$ are the probabilities amplitudes
to find the electron at the site $j$ and at the $l$-th QD
respectively, with energy  $\omega=-2t\cos k$. As we study the
paramagnetic case the spin index is neglected.

The amplitudes $a_{j}^{k}$ and $b_{l}^{k}$ obey the following linear
difference equations

\begin{subequations}
\label{e:diferencias}%
\begin{align}
\omega a_{j}^{k}  &  =-t(a_{j+1}^{k}+a_{j-1}^{k})\,,\quad j\neq0\,,\\
\omega a_{0}^{k}  & =-t(a_{1}^{k}+a_{-1}^{k})-\tilde{t}_{1}b_{1}^{k}-\tilde{t}_{2}b_{2}^{k}\,,\\
(\omega-\tilde{\varepsilon}_{1})b_{1}^{k}  &  =-\tilde{t}_{1}a_{0}^{k}\,,\\
(\omega-\tilde{\varepsilon}_{2})b_{2}^{k}  &  =-\tilde{t}_{2}a_{0}^{k}\,.
\end{align}
\end{subequations}

In order to study the solutions of Eqs.~\eqref{e:diferencias}, we
assume that the electrons are described by a unitary incident
amplitude plane wave and reflection and transmission amplitudes
$r$ and $\tau$, respectively. That is,

\begin{subequations}
\label{e:solut}%
\begin{align}
a_{j}^{k}  &  = \text{e}^{\text{i}k\cdot j} + r\text{e}^{-\text{i}k\cdot j}
                \, ; \,  \,\left(  k\cdot j<0 \right)   \, ,\\
a_{j}^{k}  &  = \tau\text{e}^{\text{i}k\cdot j}
                \qquad \, ; \qquad \left(  k\cdot j>0 \right) .
\end{align}
\end{subequations}

Inserting Eqs.~\eqref{e:solut} into Eqs.~\eqref{e:diferencias}, we
get an inhomogeneous system of linear equations for $\tau$, $r$,
$a_{j}^{k}$ and $b_{l}^{k}$, leading to the following expression in equilibrium

\begin{equation}
\tau=\frac{1}{1+i ( \frac{ \tilde{\Gamma}_{1} }{ \omega-\tilde{\varepsilon}_{1} }
                   +\frac{ \tilde{\Gamma}_{2}}{ \omega-\tilde{\varepsilon}_{2} }  ) }
\,, \label{t-amplitude}%
\end{equation}

\noindent where
$\tilde{\Gamma}_{l}=\pi\tilde{t}_{l}^{2}\rho_{0}(\omega)$ ($l=1,2$)
is the renormalized coupling between each quantum-dot and the leads
of density of states $\rho_{0}(\omega)$. In spite of the apparent
simplicity of the expression, it is necessary remember that the
quantity $\tilde{t}_{l}$ implicitly depends on the expectation
values of the boson operators also as fermion operators.

The transmission probability is given by $T=|\tau|^{2}$,

\begin{equation}
T(\omega)=\frac{1}{1+ ( \frac{\tilde{\Gamma}_{1}}{\omega-\tilde{\varepsilon}_{1}}
                       +\frac{\tilde{\Gamma}_{2}}{\omega-\tilde{\varepsilon}_{2}}    )^{2} }
\,.
\end{equation}

From the amplitudes $b_{1}^{k}$ and $b_{2}^{k}$ we obtain the
local density of states (LDOS) at the quantum dot $l$ (with $l =
1, 2$). In equilibrium that is,

\begin{equation}
\rho_{l}(\omega)  = \frac{1}{\pi \tilde{\Gamma}_{l}} \,
\frac{(\frac{\tilde{\Gamma}_{l}}{\omega-\tilde{\varepsilon}_{l}})^2}{1+ ( \frac{\tilde{\Gamma}_{1}}{\omega-\tilde{\varepsilon}_{1}}
                       +\frac{\tilde{\Gamma}_{2}}{\omega-\tilde{\varepsilon}_{2}}    )^{2} }
\,.
\end{equation}

In the nonequilibrium case, we suppose a finite source-drain biased with a symmetric
voltage drop. The incident electrons from the left side (L), they are in equilibrium with thermodynamical potential $\mu_L =V/2$, and the incidents from the right side (R), they are in equilibrium with thermodynamical potential $\mu_R = - V/2$.

Once the amplitudes $a_{j,\sigma}^{k}$ and $b_{j,\sigma}^{k}$ are known, the electronic expectation values is obtained from,%

\begin{equation}
\langle f_{l}^{\dag } c_{j } \rangle =
\frac{1}{2} \sum_{\alpha = L,R} \frac{1}{N} \sum_{k_{\alpha}}
    f\left( \epsilon_{k_{\alpha}} -\mu_{\alpha} \right)
    b_{l}^{k_{\alpha} \ast} a_{j}^{k_{\alpha}}
\end{equation}

And the current is obtained from,
\begin{equation}
J = 2\frac{2e}{\hbar}t
   \sum\limits_{\alpha,k_\alpha } f\left( \epsilon_{k_{\alpha}} -\mu_{\alpha} \right)
          {\operatorname{Im}}\{a_{0}^{k_\alpha \ast} a_{1}^{k_\alpha}\}
\end{equation}

\noindent where $f\left( \epsilon_{k_{\alpha}} -\mu_{\alpha} \right) $ it is
the Fermi function for incident electrons from the $\alpha $ side.


\section{Results}

We solve numerically the set of nonlinear equations
and take typical values for the parameters that define the system,
$t=25\,\Gamma$, $t_{0}=5\,\Gamma$ where $\Gamma=\pi t_{0}^{2}\rho_{0}(0)$
is taken to be the unit of energy.

We consider first the situation in equilibrium where the two dots
local state energies are set by $\varepsilon_{1}=V_{g}-\delta
V$, and $\varepsilon _{2}=V_{g}+\delta V$. We choose the value
of $V_{g}=-3\Gamma$. From now on all energies in units of $\Gamma$.

\begin{figure}[h]
\vspace{20pt}
\centering
\includegraphics[angle=0, scale=0.5]{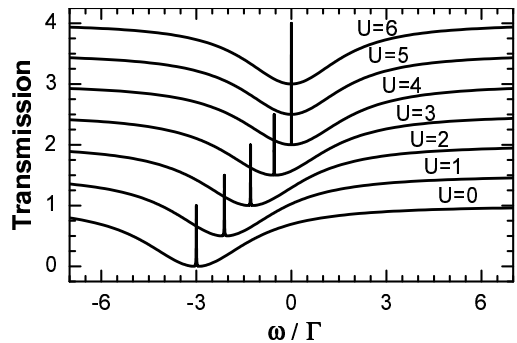}
\caption{Transmission spectrum in equilibrium for $V_{g}= -3$,
         $\delta V = 0.1 $ and various values of $U$.}
\label{f:fig2}
\end{figure}

The transmission probability, $T$, is displayed in Fig.~\ref{f:fig2}
for various values of $U$. The transmission probability always
reaches zero at $\omega=\tilde{\varepsilon}_{1}$ and
$\tilde{\varepsilon}_{2}$ and unitary value at
$\omega=(\tilde{\varepsilon }_{1}+\tilde{\varepsilon}_{2})/2$. For
small values  of $U$ the anti-ferromagnetic spin-spin correlation
between the dots is dominant and the system does not posses a Kondo
regime\cite{enrique}. Increasing $U$, a sharp feature develops close
to the Fermi energy ($\omega=0$), indicating the appearance of a
Kondo resonance.

For $U$ sufficiently large the transmission can be written approximately as
the superposition of a Fano and a Briet-Wigner line shapes,
\begin{equation}
T(\omega)\approx\frac{(\epsilon
+q)^{2}}{\epsilon^{2}+1}+\frac{{\tilde{\Delta}}^{2}}{\omega^{2}+{\tilde{\Delta}}^{2}}\,,
\label{e:transmission}
\end{equation}
where $\epsilon=\omega/2\tilde{\Gamma}$, $q=0$, with
${\tilde{\Delta}}=\delta\tilde{V}^{2}/2\tilde{\Gamma}$.

\begin{figure}[h]
\vspace{20pt}
\includegraphics[angle=0, scale=0.5]{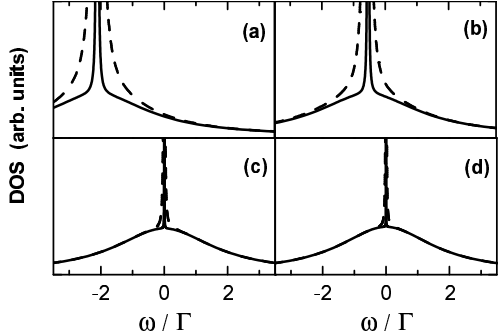}
\caption{DOS for $V_{g}=-3$ ,  $\delta V = 0.1$ (solid line),
     0.5 (dashed line). The on site energy $U$, is
    (a) 1, (b) 3, (c) 5 and (d) 6. }
\label{f:fig3}%
\end{figure}

The DOS gives us more details about the formation of the Kondo
resonance. The DOS is displayed in Fig.~\ref{f:fig3}.  In the
Kondo regime the DOS can be written as the superposition of the
two Lorentzian. These results imply the existence of two Kondo
temperature $T_{1K}=2\tilde{\Gamma}$ and
$T_{2K}=\tilde{\Delta}=\tilde{\delta V}^{2}/\tilde{\Gamma}$,
associated to each Kondo regime.

\begin{equation}
\rho(\omega)\approx\frac{1}{\pi}\frac{2\tilde{\Gamma}}{\omega^{2}%
+4\tilde{\Gamma}^{2}}+\frac{1}{\pi}\frac{\tilde{\Delta}}{\omega^{2}%
+\tilde{\Delta}^{2}}\,.
\end{equation}

The above behavior of the DOS is due to quantum interference taking
place in the transmission through the two different discrete states
(the two quantum-dot levels) coupled to common leads. This
phenomenon resembles the Dicke effect in optics, which takes place
in the spontaneous emission of a pair of atoms radiating a photon
with a wave length much larger than the separation between
them.~\cite{dicke}
    The luminescence spectrum is
characterized by a narrow and a broad peak, associated with long
and short-lived states, respectively. The former state, weakly
coupled to the electromagnetic field, is called \emph{subradiant},
and the latter, strongly coupled, \emph{superradiant\/} state. In
the present case this effect is due to the indirect coupling
between up-down QDs through the QW. The states strongly coupled to
the QW yield an effective width $2\tilde{\Gamma}$ while those
weakly coupled to the QW give a  Dicke state with width
$\tilde{\Delta}$.
    A similar result was found for a parallel double quantum
dot without electron-electron interaction.\cite{Orellana04}

\begin{figure}[h]
\centering
\includegraphics[angle=0, scale=0.5]{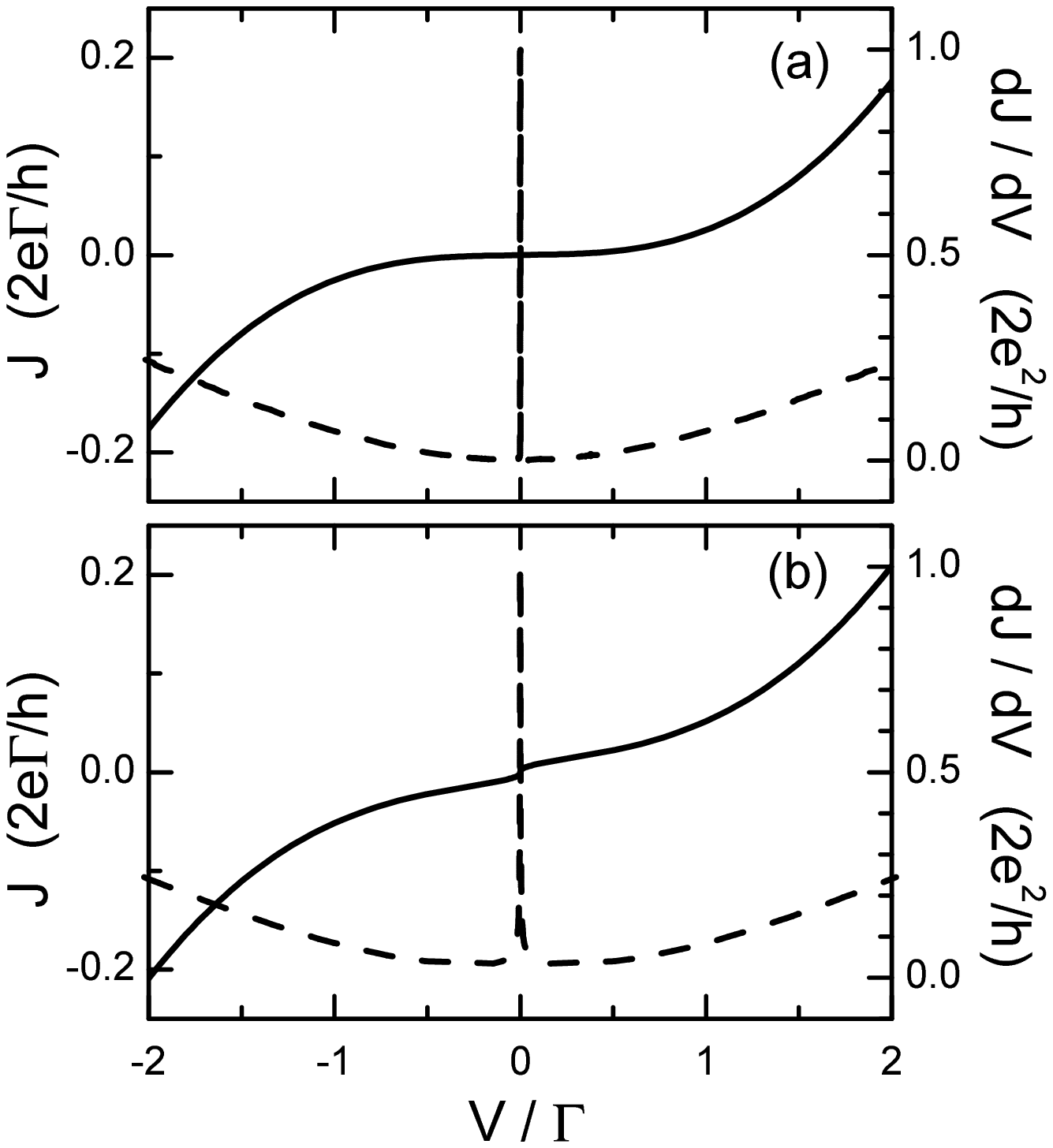}
\caption{Current (solid line) and differential conductance (dashed
line) for                 $V_{g}=-3$,  on site energy,$U=6$ for a)
$\delta V = 0.1 $ and
         b) $\delta V =0.5 $}
\label{f:fig4}
\end{figure}

The current and the differential conductance $dJ/dV$ are two
significant and experimentally measured quantities, which have
been calculated numerically at finite source-drain biases.

Figure \ref{f:fig4} displays the characteristic $J-V$ (solid line)
and the differential conductance $dJ/dV$-$V$(dashed line) for two
values of $\delta V$. For $\delta V = 0.1 \Gamma$ the current shows
a pronounced plateau around zero bias while for $\delta V = 0.5
\Gamma$ the plateau is less defined. However in both cases the
differential conductance shows an anomaly at zero bias.

\begin{figure}[h]
\centering
\includegraphics[angle=0, scale=0.5]{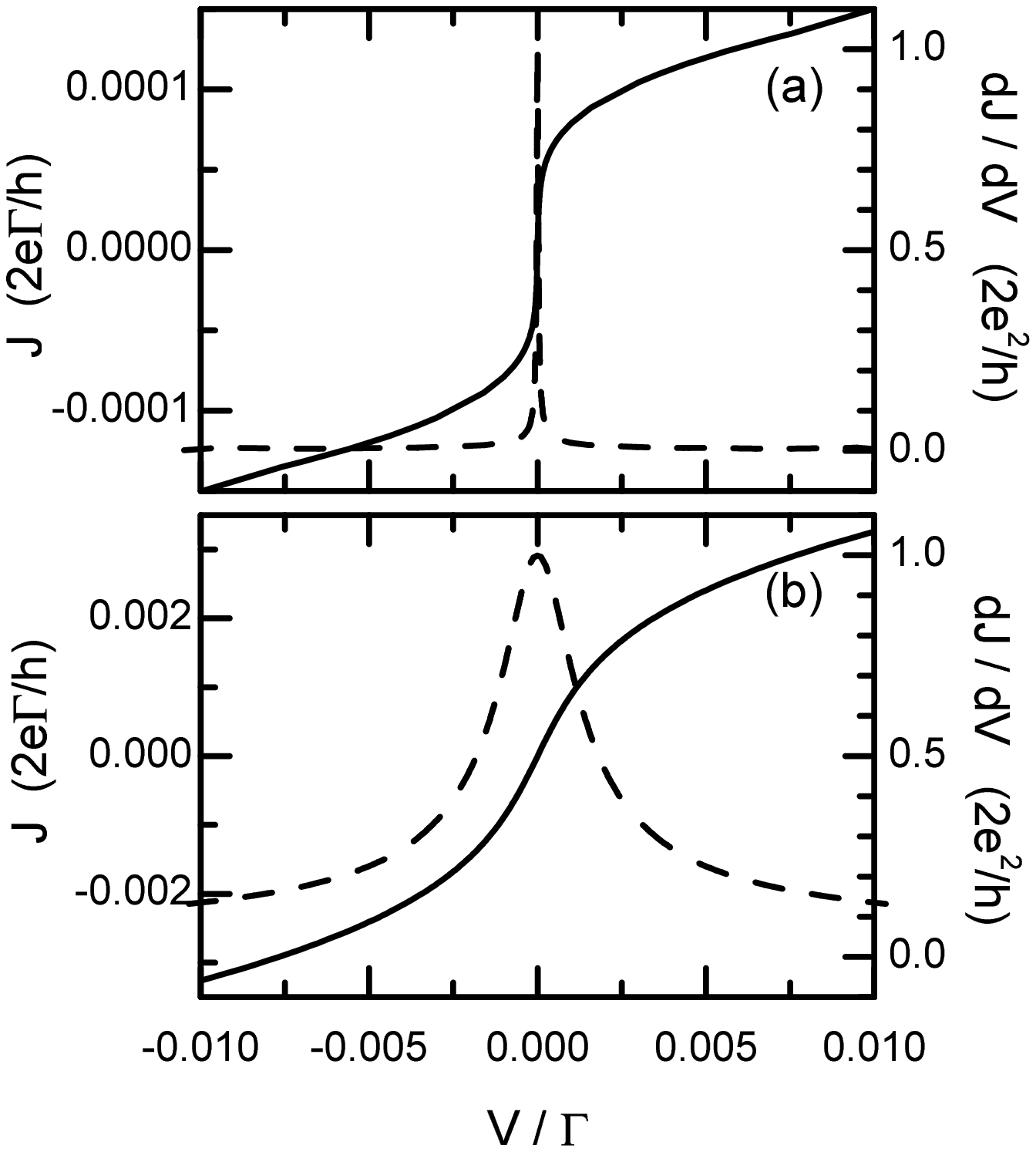}
\caption{Current (solid line) and Differential conductance (dashed line) for                   $V_{g}=-3$, on site energy, $U=6$ for a) $\delta V = 0.1$
          and b)$\delta V = 0.5$}
\label{f:fig5}
\end{figure}

Figure \ref{f:fig5} shows details of the current and differential
conductance around zero bias.

We can obtain  the expressions for the current and the differential
conductance by integrating over $\omega$ the transmission
probability given in Eq.~\eqref{e:transmission}.

\begin{eqnarray}
J&\approx&
  \frac{2e}{h} \left[ V
               - 2\tilde{\Gamma}\arctan\left( \frac{V}{2\tilde{\Gamma}} \right)
               +  \tilde{\Delta}\arctan\left( \frac{V}{\tilde{\Delta}}\right)
              \right] , \nonumber \\
\frac{\partial J}{\partial V}   & \approx&
  \frac{2e^{2}}{h} \left[ 1
                  - \frac{4\tilde{\Gamma}^{2}}{{\left( \frac{V}{2} \right) }^{2} + 4{\tilde{\Gamma}}^{2}}%
                 +\frac{{\tilde{\Delta}}^{2}}{\left( \frac{V}{2}\right) ^{2}                                + {\tilde{\Delta}}^{2}}\right ]  \,,
\label{current}
\end{eqnarray}

We identify each term of the above equation as follows. The first
term in the right side of the Eqs.\eqref{current} is the
contribution arising from an ideal unidimensional conductor. The
second term comes from the Kondo-Fano state with temperature
$T_{1k}$  giving a quasi plateau for the current and almost zero
differential conductance when $|V|\ll \tilde{\Gamma}$.  The third
term arises from the Kondo-Dicke state weakly coupled to the wire.
It is responsible for an abrupt increase of the current and an
amplification on the differential conductance around zero bias.
Finally, for $|V|>\tilde{\Gamma}$, Kondo effect disappears.

\section{Summary}

We have studied the transport through two single side-coupled
quantum dots using the finite-$U$ slave boson mean field approach
at $T=0$. We have found that the transmission spectrum shows a
structure with two anti-resonances localized at the renormalized
energies of the quantum dots. The DOS of the system shows that
when the Kondo correlations are dominant there are two Kondo
regimes each with its own Kondo temperature. The above behavior of
the DOS is due to quantum interference in the transmission through
the two different resonance states of the quantum dots coupled to
common leads. This result is analogous to the Dicke effect in
optics. These phenomena have been analyzed as a function of the
relevant parameters of the system.

\section*{Acknowledgments}

G.A.L. and P.A.O. would like to thank financial support Milenio
ICM P02-054F, P.A.O. also thanks FONDECYT (grants 1060952 and
7020269), and G.A.L. thank U.A. (PEI-1305-04). E.V.A. acknowledges
support from the brazilian agencies CNPq (CIAM project) and
FAPERJ.


\begin{thebibliography}{99}                                                                                               %
\bibitem {GR88}L.~I.~Glazman, M.~{\'E}.~Raikh, JETP Lett. {\textbf 47},  452 (1988);
T.~K.~Ng, P.~A.~Lee, Phys. Rev. Lett. {\textbf 61},  1768 (1988);
\bibitem{GSMAMK98} D.~Goldhaber-Gordon, H.~Shtrikman, D.~Mahalu, D.~Abusch-Magder, U.~Meirav,
M.~A.~Kastner, Nature {\textbf 391},  156 (1998);
D.~Goldhaber-Gordon, J.~G{\"o}res, M.~A.~Kastner, H.~Shtrikman,
D.~Mahalu, U.~Meirav, Phys. Rev. Lett. {\textbf 81},  5225 (1998).

\bibitem{Kouwenhoven}S.~M.~Cronenwett, T.~H.~Oosterkamp, L.~P.~Kouwenhoven, Science
{\textbf 281},  540 (1998).

\bibitem{THCP02} R. Franco, M. F. Figueira, E. V. Anda, Phy. Rev. B {textbf 67} 155301 (2003);
 M.~E.~Torio, K.~Hallberg, A.~H.~Ceccatto, C.~R.~Proetto, Phys. Rev. B {\textbf 65}, 085302 (2002).

\bibitem {GGHK00}J.~G{\"o}res, D. Goldhaber Gordon, S. Heemeyer,
M. A. Kastner, Phys. Rev. B {\textbf 62}, 2188 (2000).

\bibitem{fano} U. Fano, Phys. Rev. \textbf{124}, 1866 (1961).


\bibitem {DRVRPM00}P. Debray, O. E. Raichev, P. Vasilopoulos, M. Rahman,
R. Perrin, W. C. Mitchell, Phys. Rev. B {\textbf 61}  10950
(2000).

\bibitem {KCKS01}K.~Kang, S. Y. Cho, J. J. Kim, S.-C. Shin, Phys. Rev. B {\textbf
63},  113304 (2001).

\bibitem{tanaka} Yoichi Tanaka and Norio Kawakami Phys. Rev. B {\textbf 72}, 085304
(2005).

\bibitem{sakano} Rui Sakano and Norio Kawakami Phys. Rev. B {\textbf 72},  085303 (2005).

\bibitem{stefanski} Priotr Stefa\'nski, Sol. Stat. Comm.
\textbf{128}, 29 (2006).

\bibitem{glazman} Hiroyuki Tamura and Leonid Glazman, Phys. Rev. B
\textbf{72}, 121308(R) (2005).

\bibitem {KR86}G. Kotliar, A. E. Ruckenstein, Phys. Rev. Lett. {\textbf 57}  1362 (1986),
and references cited therein.

\bibitem {DL01}B. Dong, X. L. Lei, Phys. Rev. B {\textbf 63},  235306 (2001).

\bibitem {enrique}C. A. Busser, E. V. Anda, A. L. Lima, M. A. Davidovich, Phy.Rev. B {\textbf 62} 9907
(2000).

\bibitem{dicke} R. H. Dicke, Phys. Rev. {\textbf 89}, 472 (1953).

\bibitem {brandes} T. Brandes, Phys. \textbf{408}, 315 (2005).


\bibitem {C84}P. Coleman, Phys. Rev. B \textbf{29},  3035 (1984).

\bibitem {LOA03}G. A. Lara, P. A. Orellana, E. V. Anda, Solid State Comm. {\textbf
125}, 165 (2003).

\bibitem{Orellana04} P.\ A.\ Orellana, M.\ L.\ Ladr\'{o}n de Guevara, and
        F.\ Claro, Phys.\ Rev.\ B {\textbf 70}, 233315 (2005).

\end{thebibliography}
\end{document}